\documentclass[twocolumn]{aastex631}

\usepackage{newtxtext,newtxmath}
\usepackage{graphicx}	
\usepackage{xcolor}
\usepackage{anyfontsize}

\shorttitle{Radio emission from HR\,5907}
\shortauthors{Biswas et al.}

\begin{document}


\title{A Non-Stop Aurora? The Intriguing Radio Emission from the 
 Rapidly Rotating Magnetic Massive Star HR\,5907}

\correspondingauthor{Ayan Biswas}
\email{ayan.biswas@queensu.ca}

\author[0000-0002-1741-6286]{Ayan Biswas}
\affiliation{Department of Physics, Engineering Physics \& Astronomy, Queen’s University, Kingston, Ontario K7L 3N6, Canada}
\affiliation{Department of Physics \& Space Science, Royal Military College of Canada, PO Box 17000, Station Forces, Kingston, ON K7K 7B4, Canada}

\author[0000-0001-8704-1822]{Barnali Das}
\affiliation{CSIRO, Space and Astronomy, PO Box 1130, Bentley, WA 6151, Australia}

\author[0000-0003-2088-0706]{James A. Barron}
\affiliation{Department of Physics, Engineering Physics \& Astronomy, Queen’s University, Kingston, Ontario K7L 3N6, Canada}

\author[0000-0002-1854-0131]{Gregg A. Wade}
\affiliation{Department of Physics \& Space Science, Royal Military College of Canada, PO Box 17000, Station Forces, Kingston, ON K7K 7B4, Canada}
\affiliation{Department of Physics, Engineering Physics \& Astronomy, Queen’s University, Kingston, Ontario K7L 3N6, Canada}

\author[0000-0002-9296-8259]{Gonzalo Holgado}
\affiliation{Instituto de Astrofísica de Canarias, E-38200 La Laguna, Tenerife, Spain}


\begin{abstract}
HR\,5907 (HD\,142184) stands out among magnetic OB stars for its rapid rotation, exceptionally hard X-ray emission, and strong magnetic field. High-frequency ($>$5 GHz) radio emission from the star exhibits an approximately flat spectrum that can be attributed to gyrosynchrotron emission from a dense centrifugal magnetosphere. In a survey of radio emission from massive stars at sub-GHz frequencies, we noticed remarkable low-frequency radio emission from this star, characterized by high circular polarization and brightness temperature, which is inconsistent with the gyrosynchrotron model.  We present a follow-up low-frequency radio study of this star with the upgraded Giant Metrewave Radio Telescope (uGMRT) in search of emission mechanisms that can go undiagnosed at higher frequencies. We detect variable radio emission characterized by varying degrees of circular polarization (15--45\%) throughout the rotation cycle. The broad-band spectral fitting also suggests additional emission components at lower frequencies. We show that the observed emission is likely auroral emission via electron cyclotron maser emission (ECME), and identify this star as a Main-sequence Radio Pulse emitter (MRP). For MRPs, ECME is usually observed as short polarized enhancements near the magnetic nulls of the star. The detection of a high degree of circular polarization ($>15$\%) at all times makes HR\,5907 unique among MRPs. This is only the second MRP after $\rho$ Oph C (detected polarization fraction: 0--60\%) that exhibits persistent coherent radio emission attributed to the nearly aligned stellar magnetic and rotational axes.

\end{abstract}


\section{Introduction} \label{sec:intro}

About 1 in 10 hot stars are found to be magnetic, possessing strong, predominantly dipolar magnetic fields \citep{Wade2014, Morel2015, Grunhut2017}. The presence of these strong magnetic fields can create magnetically confined winds (MCWs), creating a co-rotating magnetosphere surrounding the star \citep{Babel1997}. The MCW is proposed to be responsible for the reduction of rotational angular momentum, resulting generally in lower projected rotational velocities with respect to non-magnetic counterparts \citep{Donati2009, Shultz2018}. Among magnetic hot stars, younger stars are found to have shorter rotation periods, whereas older stars show systematically longer rotation periods \citep{Shultz2017, Erba2021}.

The presence of a MCW can lead to magnetospheric emission at different wavelengths (e.g. H$\alpha$, \citealt{Petit2013,Shultz2020}; radio, \citealt{Linsky1992, Leto2021, Shultz2022}; X-rays, \citealt{Naze2014}; UV, \cite{Erba2021}; infrared, \citealt{Oksala2015}). In radio bands, emission from a stellar magnetosphere is dominated by the gyrosynchrotron mechanism that can be produced via the centrifugal breakout (CBO) mechanism \citep{Shultz2022, Owocki2022}. This scenario prefers stars with rapid rotation and strong magnetic fields. In rapid rotators, centrifugally supported plasma above the Kepler co-rotation radius ($R_K$) forms a centrifugal magnetosphere (CM). The plasma in the CM is ejected via CBO when an instability threshold is reached, producing magnetic reconnection that can fuel radio emission. In a broad-band radio study of hot stars, \cite{Shultz2022} show that young rapid rotators with CMs are indeed radio luminous compared to old systems lacking CMs where radio emission is generally not detected.

In addition to non-thermal gyrosynchrotron emission, some magnetic B stars also show coherent emission at low radio frequencies in the form of highly circularly polarized radio pulses. The repeating pulses are usually found to occur at phases near the magnetic null of the star (i.e. when the longitudinal magnetic field $\langle B_\mathrm{z}\rangle$ is zero). The emission mechanism for this phenomenon is now broadly believed to be electron cyclotron maser emission (ECME, \citealt{Trigilio2000}), and the phenomenon has been detected in many magnetic B-type stars with appropriate magnetic geometries (e.g. \citealt{Trigilio2000, Leto2020a, Leto2020b, Das2019a, Das2020b, Das2022b}).

\subsection[The Target: HR 5907]{The Target: HR\,5907} \label{tab:parameters}

HR\,5907 (HD\,142184) is a main sequence B2.5V star \citep{Hoffleit1991}, standing out among other magnetic OB stars for its rapid rotation, exceptionally hard X-ray emission and strong magnetic field. From a study of its photometric and spectroscopic variability, \cite{Grunhut2012_hr5907} found a rotation period of $\sim 0.51$ days, making it one of the most rapidly rotating magnetic early B-type stars known to date. The fast rotation makes the star oblate ($R_{\rm polar}/R_{\rm eq} \approx 0.88$). The inclination of the star's rotation axis to earth's line-of-sight is found to be $\sim 70^{\circ}$. To date, no clear evidence of binarity has been reported for this star.

An additional remarkable property of HR\,5907 concerns its strong magnetic field. \cite{Grunhut2012_hr5907} modeled the longitudinal magnetic curve of HR\,5907 assuming a simple dipole of polar magnetic field strength $B_{\rm p}$ = 15.7 kG, with a magnetic obliquity of $\beta \approx 7^{\circ}$. On the other hand, the authors reported a lower polar field strength of 10.4 kG from the Bayesian analysis of the optical Stokes V profiles. \cite{Grunhut2012} suggested that this disagreement may be due to a more complex topology of the magnetic structure, or to a non-uniform distribution of helium on the star’s surface. A summary of the stellar and magnetospheric parameters is given in the Appendix \ref{stellar_parameters}. The rapid rotation and strong magnetic field thus make this star an ideal target for studying magnetospheric emission in an extreme environment.

\subsection{Previous Radio Observations} \label{sec:previous}

\begin{figure}  
    \begin{center}
    \includegraphics[width=0.46\textwidth]{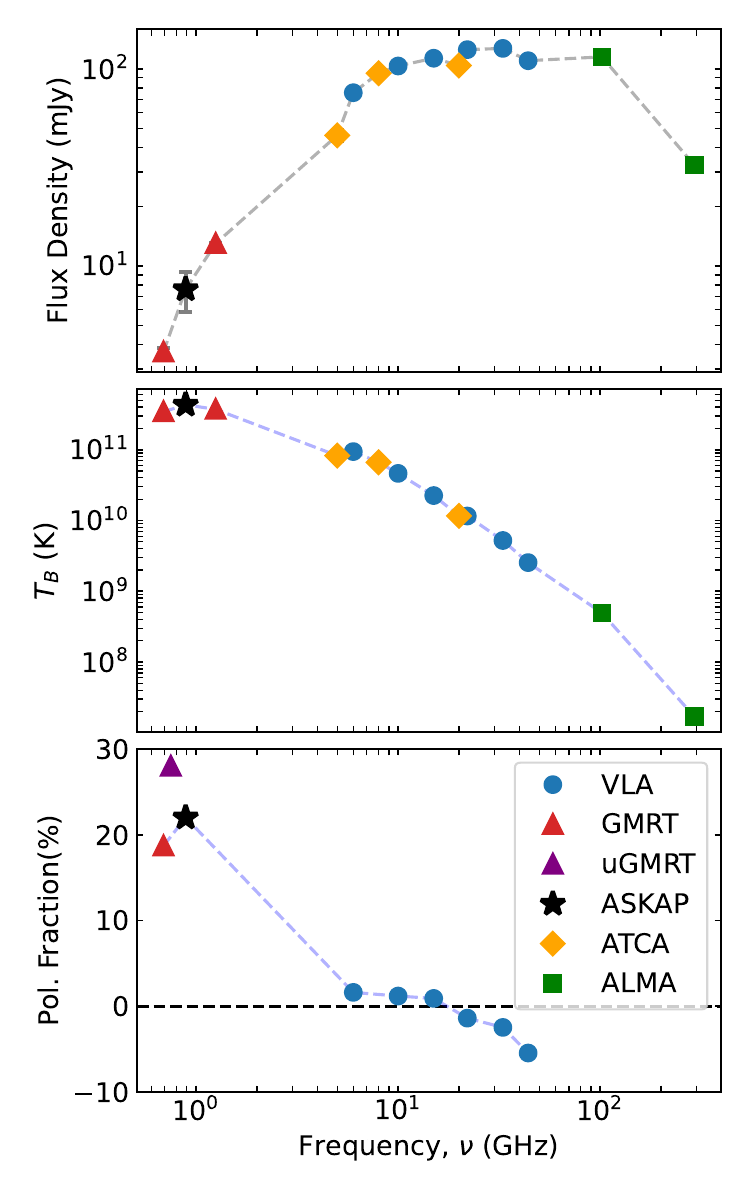}  
        \caption{(Top) Radio spectra of HR\,5907 combining different telescopes. See text (Sec. \ref{sec:previous}) for details on various data points \citep{Murphy2010, Leto2018, Pritchard2021}. (Middle) Brightness temperature in each case, assuming the whole stellar surface as the emitting area. (Bottom) Fraction of circular polarization calculated from Stokes V images. GMRT and uGMRT results are from this work. For uGMRT, observed mean circular polarization is plotted.}
        \label{fig:spectra}
    \end{center}
\end{figure}

The star was previously detected across a wide range of radio frequencies, from cm to mm wavelengths \citep{Leto2018}. It was observed with the Australian Telescope Compact Array (ATCA) as part of the  AT20G survey \citep{Murphy2010}. The ATCA flux densities are: 46 $\pm$ 3 mJy at 5 GHz, 95 $\pm$ 5 mJy at 8 GHz, and 104 $\pm$ 6 mJy at 20 GHz, respectively. \cite{Leto2018} observed the star with the Very Large Array (VLA) and Atacama Large Millimeter Array (ALMA) covering frequency bands from 6 GHz to 292 GHz. Recently, the star was also detected at 0.89 GHz with the Australian SKA Pathfinder (ASKAP) showing $+$22\% circular polarization \citep{Pritchard2021}. The radio spectra from these observations are shown in the top panel of Fig. \ref{fig:spectra}.

The Stokes V spectra obtained with VLA observations reveal two peculiar properties shown by the star. The first peculiarity is the increase of  the circular polarization fraction with decreasing frequency. This behaviour is unusual for similar systems (e.g. see Fig. 2 of \citealt{Leto2017} showing a decreasing trend in polarization fraction with decreasing frequency). Another peculiarity is the reversal of the sign of circular polarization with frequency. Consistent negative values of circular polarization are observed at 44 GHz, while 6 GHz observations consistently revealed positive values of the Stokes V flux. \cite{Leto2018} considered an {\it ad hoc} dipole flip assumption to match their simulations with this peculiar feature.

In this work, we report low-frequency follow-up of HR\,5907. The paper is structured as follows: Section \ref{observation} gives a brief description of observations and data analysis procedures. In Section \ref{radio}, we report the results from our observation and discuss the origin of low-frequency radio emission. We finally summarize our findings in Sec. \ref{conclusion}.

\section{Observation} \label{observation}

\subsection{GMRT Survey}

We observed the target in a survey observation carried out during cycles 27 and 28 of the legacy GMRT (Biswas et al., in prep.). The target was detected in two bands: 610 MHz and 1390 MHz. The flux densities were $3.66 \pm 0.19$ mJy and $12.99 \pm 0.12$ mJy  at 610 MHz and 1390 MHz bands, respectively. The target was also detected in the Stokes V image at 610 MHz with a flux density of 0.685 $\pm$ 0.096 mJy giving a circular polarization fraction of 18.7\%. If we assume the area of emission has the same radius as that of the star, we found the brightness temperature to be $3.43 \times 10^{11}$ and $3.72 \times 10^{11}$ K, respectively, for the 610 MHz and 1390 MHz bands. The GMRT observations are also plotted in Fig.  \ref{fig:spectra}. The circular polarization fraction and brightness temperature is found to be strikingly higher in low frequency observations compared to high frequency radio observations. We  follow up HR\,5907 with the upgraded Giant Metrewave Radio Telescope \citep[uGMRT,][]{Gupta2017} in order to determine the radio emission mechanism, and search for the possible reason for the star's  strikingly different behavior at low frequencies.

\subsection{uGMRT Observations}

To study the rotational modulation, we observed HR\,5907 for a full rotational cycle. The short rotation period of the star made it possible to observe the star using a reasonable amount of telescope time. We divided the observations into two equal-duration windows of 7 hours each. The observations were carried out on 1 and 15 August 2024 during uGMRT Cycle 46   (ObsID: 46\textunderscore 065, PI: A. Biswas). During the first observation, we observed in dual sub-array mode, where we performed simultaneous measurements in Band 4 (550--900 MHz) and Band 5 (1050--1450 MHz) of the uGMRT. The sub-array mode observation was aimed to measure the spectral index reliably at each rotational phase. However, during the second observing run, we observed only in Band 4 in full-array configuration to achieve better sensitivity. In both cases, the bands were divided into 2048 channels. We used 3C286 (J1331+305) as the amplitude and band-pass calibrator, and J1626-298 was used as the phase calibrator.

The data analysis was performed using the Common Astronomy Software Applications (CASA\footnote{\url{https://casa.nrao.edu/}}) package \citep{McMullin2007}. The calibration and imaging process is similar to the analysis of \cite{Biswas2023}. The final bandwidth of the calibrated data was $\sim$290 MHz for Band 4 observations, and $\sim 370$ MHz for the Band 5 observation. For the target field, the data were averaged over 4 frequency channels to reduce computational cost, continuum images were obtained using the CASA task {`\textit{tclean}'} along with the {`\textit{mtmfs}'} deconvolver (Multi-term Multi-frequency with W-projection, \citealt{Rau2011}) and Briggs weighting. Several rounds of phase and amplitude self-calibration were performed, improving image quality. Finally, strong background sources were removed from the field using task {`\textit{uvsub}'}, and snapshot images were obtained in order to study variability.

\subsection{Ephemeris} \label{Ephemeris}

Due to HR\,5907's fast rotation rate, small uncertainties in the period can lead to large phase uncertainties over time. Additionally, the stellar rotation period might change measurably over long time spans due to magnetic braking. To help set phase constraints on the GMRT observations, we performed a period analysis on archival \textit{Hipparcos} photometry, \textit{Gaia} $G$-band photometry and EW measurements from newly obtained spectra from the robotic 1-m Hertzsprung SONG telescope. Details of the photometric and spectroscopic observations along with the period analysis process are discussed in Appendix \ref{Ephemeris_App}. We found no significant change in the rotation period over the $\sim 22$ yr time span between the \textit{Hipparcos} and \textit{Gaia} observations. To phase the uGMRT observations we adopt the ephemeris obtained from SONG H$\alpha$ EW measurements, with fixed period 0.5083 d, and $\mathrm{BJD}_{0}^{\mathrm{SONG}}=2460443.684(27)$. We estimate a phase uncertainty of $\sigma_{\phi}=0.054$ of the uGMRT observations based on the current ephemeris. 

As the uGMRT observations were obtained approximately 14 years after the magnetic measurements reported by \cite{Grunhut2012_hr5907} we are unable to accurately phase both data sets with a single ephemeris. \cite{Grunhut2012_hr5907} showed the H$\alpha$ EW maximum corresponds to the maximum longitudinal magnetic field strength ($\langle B_{z}\rangle$). Therefore, in the interpretation of our results we assume that rotation phase $\phi=0$ given by the SONG ephemeris corresponds to the maximum of the $\langle B_{z}\rangle$ curve.

\begin{figure}
    \centering
    \includegraphics[width=0.96\linewidth]{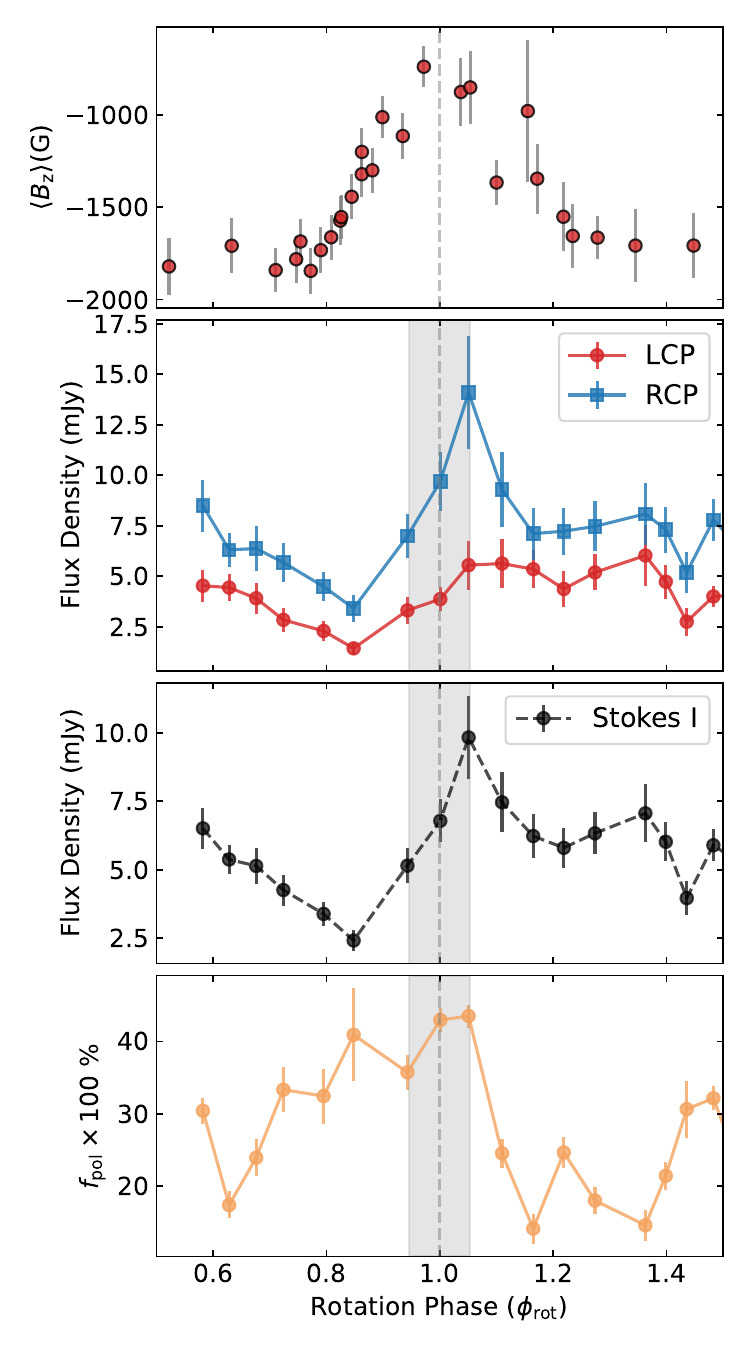}
    \caption{(a) Variation of observed longitudinal magnetic field of HR\,5907 by \cite{Grunhut2012_hr5907}. The $\langle B_{z}\rangle$ measurements are phased according to the ephemeris provided by  \cite{Grunhut2012_hr5907}. (b) light curves for right circularly polarized (RCP) and left circularly polarized (LCP) flux densities at uGMRT band 4 (710 MHz), showing maximum emission near phase 1.0. (c) The corresponding total intensity (Stokes I) light curve of the star, and (d) degree of circular polarization, defined as $\{(RCP-LCP)/(RCP+LCP)\} \times 100$\%. Here, phase 1.0 (vertical gray dashed line) corresponds to the minimum of H$\alpha$ emission, or the magnetic maximum and the gray band represents the $1 \sigma$ phase uncertainty from the SONG ephemeris. The uGMRT observations are phased according to the ephemeris provided in Appendix~\ref{Ephemeris_App}. For all observations, additional 10\% absolute flux density calibration uncertainties were added. During phases $\sim$1.25--1.40 where Band 4 flux densities are available from both observations, an average value is plotted. }
    \label{fig:band4_radio}
\end{figure}

\section{Results} \label{radio}

\begin{figure}
    \centering
    \includegraphics[width=0.44\textwidth]{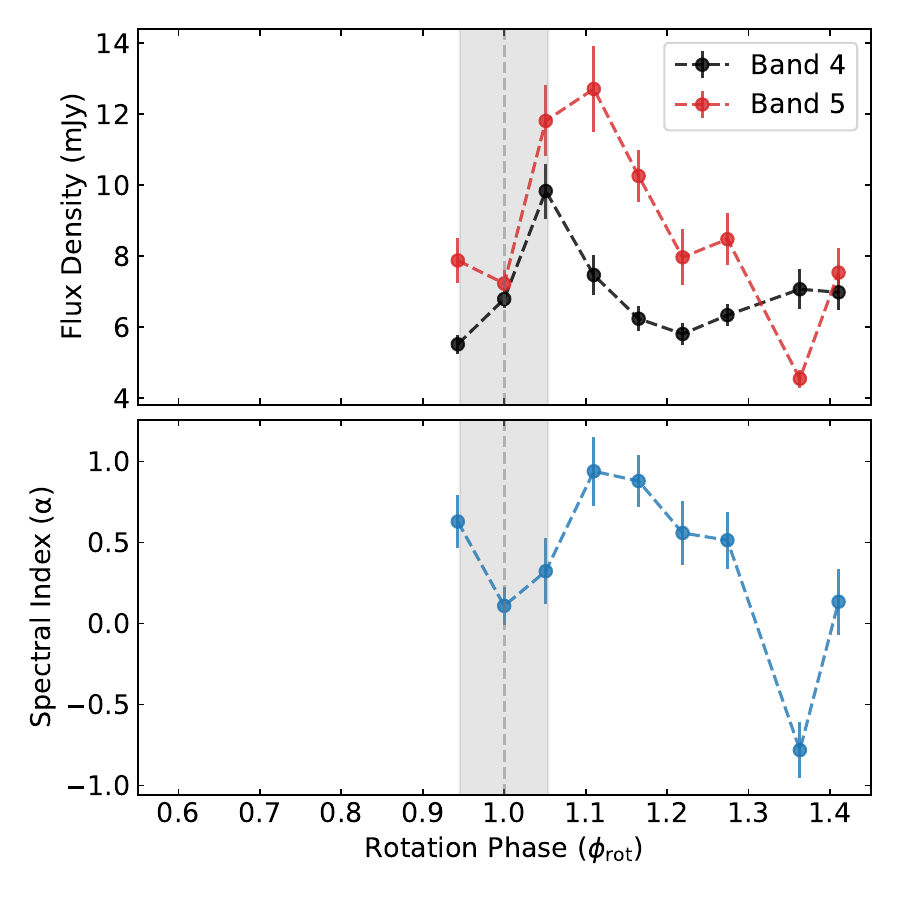}
    \caption{The top panel shows the comparison of the Band 5 (red) and Band 4 (black) light curves, and the bottom panel shows the variation of spectral index, $\alpha$ with rotation phase during the first simultaneous sub-array observation. Similar to Fig. \ref{fig:band4_radio}, phase 1.0 refers to magnetic maximum, maximum, and the gray band represents the phase uncertainty.}
    \label{fig:band5}
\end{figure}

The target was detected at all epochs. For each scan, right circularly polarized (RCP) and left circularly polarized (LCP) emission  were imaged separately in search of ECME\footnote{We have followed the IAU/IEEE convention for circular polarization}. The Band 4 (final central frequency 710 MHz) radio light curve phased according to the rotation period is presented in Fig. \ref{fig:band4_radio}. The top panel shows the comparison between RCP and LCP light curves, the middle panel shows the average Stokes I flux density, and the bottom panel shows the degree of circular polarization.  The target remains extremely bright throughout the rotation cycle. For all observations, additional 10\% flux density calibration uncertainties were added to the original error. As the observation during the first epoch ($\phi_{\rm rot} = $ 0.85 to 1.35) was taken in sub-array mode with half the number of antennae, the corresponding errors in flux densities during this period remain higher than the second observation. In Fig. \ref{fig:band5}, we show the phased light curve from Band 5 data, and the calculated inter-band spectral index variation with rotation phase.

The Band 4 Stokes I phased light curve shows a clearly non-sinusoidal nature, with maximum flux near magnetic maximum ($\phi_{\rm rot} = 0.0$ or 1.0). The total flux is dominated by RCP, which always remained higher, resulting in positive circular polarization throughout the rotation timescale. The average polarization fraction is around $+28$\%. However, for Band 5 observations, circular polarization information could not be obtained as the observations were taken with a linear feed. The Band 5 light curve also shows a variable nature, with maximum flux near phase 0.05. The spectral index was found to vary significantly, ranging between $-1$ and $+1$.

The average Stokes I flux density in Band 4 is $\sim5.7$ mJy.  The GMRT 610 MHz and 1490 MHz observations were taken at phases 0.29 and 0.44, approximately 9 years before the recent observations. The inter-band spectral index ($\alpha$) was found to be $1.54 \pm 0.06$. Then we can predict the flux density at 710 MHz to be $4.62 \pm 0.24$ mJy. This value is significantly smaller than the flux density observed at similar phases in the new observations, suggesting either a long-term change of the system, or the presence of strong cycle-to-cycle variability. However, the percentage of circular polarization remains the same within errors (polarization fraction $\sim19$\% for legacy GMRT observation). It should be noted that while calculating this spectral index, we assume that the stellar flux was not significantly different at these two phases, which is not suggested from the new light curve. As the spectral index measurements in Fig. \ref{fig:band5}b were obtained from simultaneous observations, the observed significant variation between phase 0.3 and 0.4 (i.e. 1.3 and 1.4) are reliable. Thus it is impossible to confirm whether there is long term variability with the existing observations.

\section{Discussion} \label{sec:discussion}

\subsection{Emission Mechanism}

In the observed radio emission, a significant contribution from thermal emission can be easily excluded. The thermal emission can be predicted using the formalism described by \cite{Bieging1989}. Using the stellar parameters given in Appendix \ref{tab:parameters}, we predict the free-free flux density to be of the order of $10^{-5}$ mJy. This value is negligible compared to the observed emission, and the entire emission can be attributed to be of non-thermal origin. In solitary magnetic massive stars, non-thermal radio emission is generally attributed to gyrosynchrotron emission and coherent ECME produced in auroral regions. For gyrosynchrotron emission, the polarization fraction always remains comparatively low (i.e. $< 10$\%, \citealt{Leto2019, Leto2021}). 
However, in our case, we see a minimum of 15\% circular polarization at all phases at sub-GHz frequencies. Furthermore, by considering the entire star as the emitting surface, we obtained the brightness temperature ($T_B$) to be $\sim 10^{12}$ K. Based on the observed high circular polarization at sub-GHz frequencies and high brightness temperature, we propose that the underlying emission mechanism is ECME and thus our target is another `Main sequence Radio Pulse emitter' (MRP, \citealt{Das2021, Das2022scaling}). The fact that we observed RCP emission from the southern magnetic hemisphere (negative $\langle B_\mathrm{z}\rangle$) suggests that the emission is produced in the ordinary mode \citep[also observed from the MRP HD\,142301,][]{Leto2019}.

\begin{figure}
    \centering
    \includegraphics[width=0.45\textwidth]{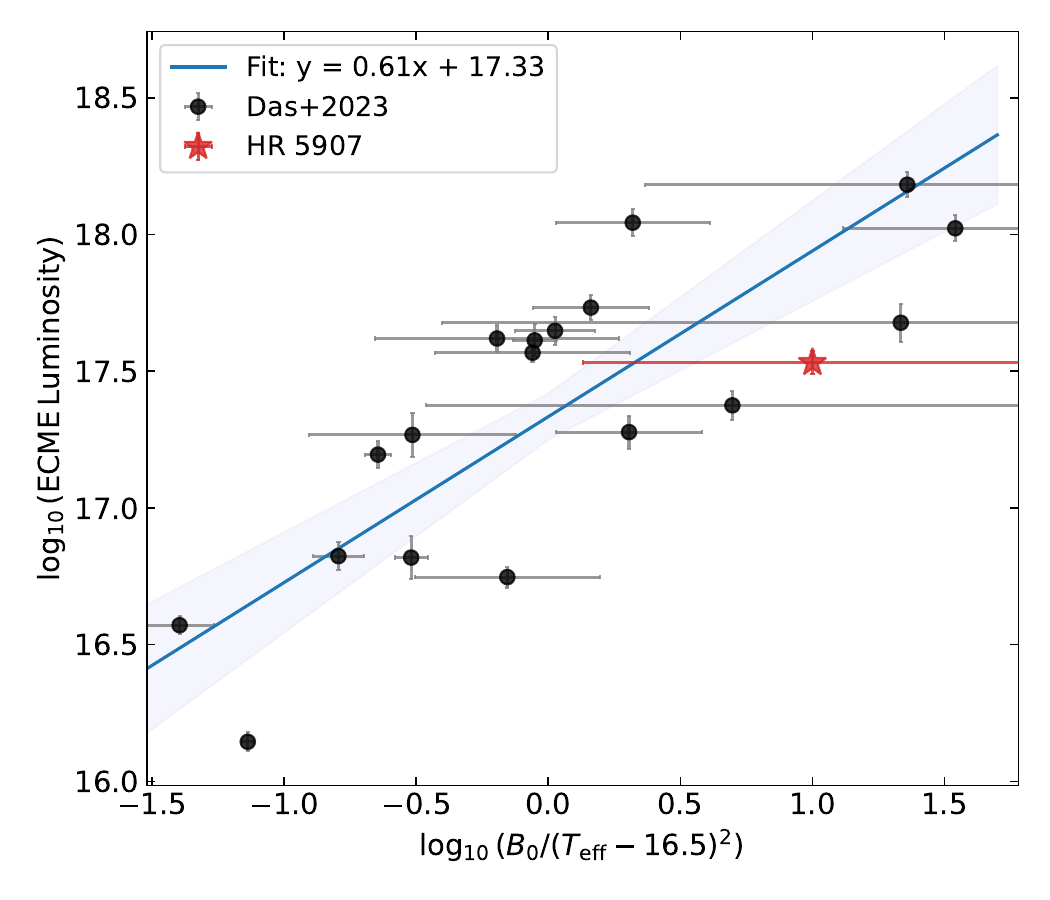}
    \caption{Position of HR\,5907 in the ECME scaling relationship diagram presented by \cite{Das2022scaling}. The black data points are from \cite{Das2022scaling}, while the position of HR\,5907 is shown in red.}
    \label{fig:ecme_scaling}
\end{figure}

We show the position of HR\,5907 in the ECME scaling diagram by \cite{Das2022scaling} in Fig. \ref{fig:ecme_scaling}. The observed luminosity falls in the scaling relation within errors. Near phase 0.0 ( where $\langle B_\mathrm{z}\rangle$ is closest to zero) the short duration of the radio emission peak ($<3$ hrs), very high circular polarization fraction ($>40$\%) and high brightness temperature ($>10^{12}$ K) makes ECME a highly likely scenario, making HR\,5907 a likely new addition to MRP family. However, no other MRPs detected till date show consistent high degree of circular polarization throughout rotation cycle, making this system unique. However, one notable system, $\rho$ Oph C, shows evidence of near-continuous auroral emission with a broad range of circular polarization fraction ($0-60$\% at 2.1 GHz, \citealt{Leto2020b}). The inclination and obliquity of $\rho$ Oph C ($i \approx 74^{\rm o}$, $0^{\rm o}<\beta<10^{\rm o}$) are comparable to those of our target, and based on the analysis by \cite{Leto2020b}, we should indeed observe auroral emission at all phases for HR\,5907. For the phases far from phase 0.0, however, it is possible that there are additional non-negligible contributions from other non-thermal emission processes. We explore this possibility in section \ref{sec_SED_fit}.

\subsection[Simulating visibility of ECME from HR 5907]{Simulating visibility of ECME from HR\,5907}\label{subsec:simulation_lc}
\begin{figure*}
    \centering
    \includegraphics[width=0.4\textwidth]{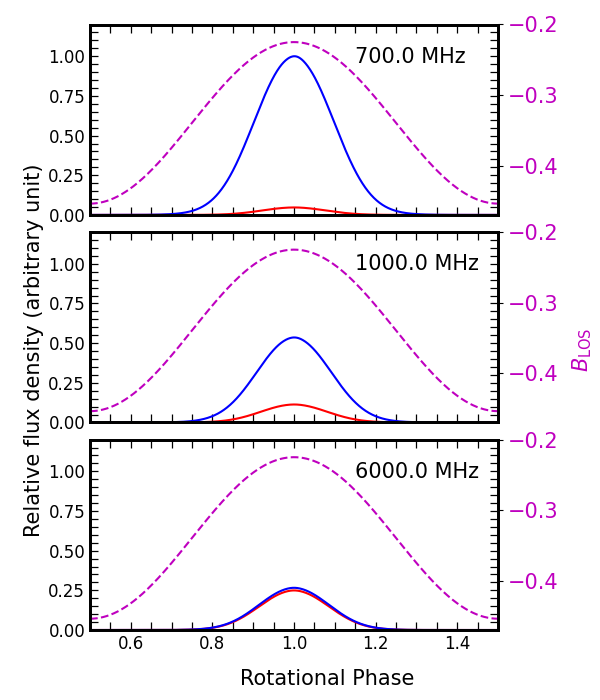}
    \includegraphics[width=0.4\textwidth]{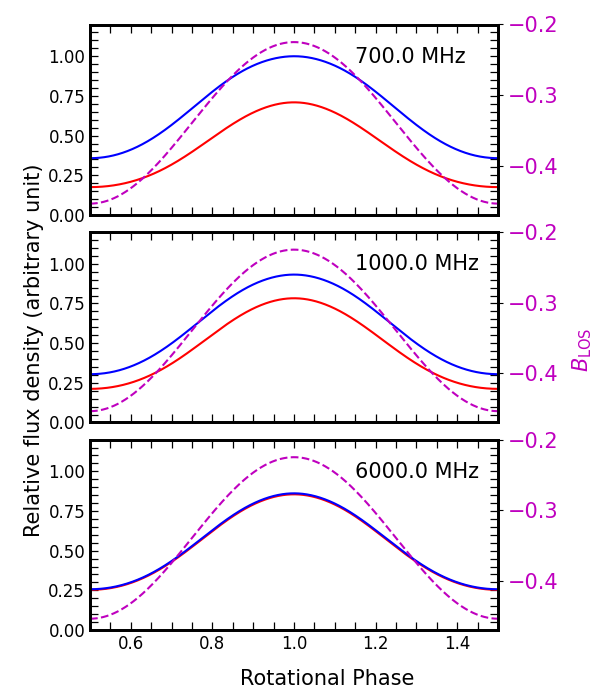}
    \caption{Simulated ECME lightcurves for a star with an axi-symmetric dipolar magnetic field, obtained using the framework of \citet{Das2020b}. For the lightcurves on the \textbf{left}, the intensity of the emission falls as a Gaussian with a mean of zero and a $\sigma$ of $5^\circ$ as the angle between the ray direction and line of sight increases. For those at \textbf{right}, we use a larger $\sigma$ of $15^\circ$. The flux densities are normalized by the maximum flux densities observed over these frequencies. The polar magnetic field strength, inclination angle and obliquity are taken to be 15.7 kG, $70^\circ$ and $7^\circ$ respectively (similar to those estimated for the target HR\,5907). The magneto-ionic mode is taken to be ordinary and the harmonic number is set to 1.
    The variation of the projection of the magnetic field vector along the line-of-sight (assuming a magnitude of unity), denoted by
     $B_\mathrm{LOS}$, is shown as the dashed magenta line, while the red and blue curves correspond to ECME produced in the northern and southern magnetic hemisphere respectively. For more details of the simulation, refer to \S\ref{subsec:simulation_lc}.}
    \label{fig:lc_simulation}
\end{figure*}

Following the tangent plane beaming model of \citet{Trigilio2011}, ECME is expected to be visible around the rotational phases where the stellar longitudinal magnetic field $\langle B_\mathrm{z}\rangle$ is zero, known as magnetic nulls. The $\langle B_\mathrm{z}\rangle$ for the star HR\,5907, however, remains consistently negative throughout the rotation cycle and thus does not exhibit magnetic nulls (see top panel of Fig. \ref{fig:band4_radio}). Nevertheless, it is possible to detect ECME from the star around the rotational phase(s) where $\langle B_\mathrm{z}\rangle$ is closest to zero. This is due to the fact that although the radiation is directed parallel to the magnetic equatorial plane, the beams suffer refraction while passing through the stellar magnetospheric plasma, so that their final direction can differ significantly from their original direction \citep{Trigilio2011,Leto2016, Das2020b}. This possibility was considered by \citet{Leto2018} to investigate whether it is possible to observe ECME from this star over the frequency range of 6--44 GHz (their observed frequencies). They concluded that such emission might be detectable at those frequencies provided the star emits ECME within a very large beaming pattern (beam-size larger than 45$^\circ$). This is in sharp contrast to the highly directed nature of ECME.

Here we examine the conditions under which we can observe ECME from this star at much lower frequencies (700 MHz to 6 GHz) under different geometrical configurations. We used the 3D framework of \citet{Das2020b} that allows continuous refraction in the stellar magnetosphere for an arbitrary density distribution. We assume an axi-symmetric dipolar magnetic field of polar strength 15.7 kG. The inclination angle and the obliquity are taken to be 70$^\circ$ and 7$^\circ$ respectively \citep{Grunhut2012_hr5907}. For such small obliquity, the magnetospheric plasma distribution is expected to be symmetric in magnetic azimuth, with a disk-like overdense region at the magnetic equator \citep[according to the `Rigidly Rotating Magnetosphere' model of][]{Townsend2005}. As ECME is produced above the magnetic poles and then travels approximately parallel to the magnetic equator, none of the beams is expected to pass through the high-density plasma disk in this case. We hence consider a simple magnetospheric density distribution of the form $n_\mathrm{p}=n_\mathrm{p0}/r$ \citep{Leto2006}, where $r$ is the radial distance in units of stellar radius $R_*$. We set $n_\mathrm{p0}=10^9\,\mathrm{cm^{-3}}$. ECME is assumed to be produced in the ordinary mode (since the enhancement is primarily observed in RCP, even though $B_\mathrm{z}$ is negative) at the fundamental mode. We also assume that the emission sites are located along magnetic field lines of equatorial radii 30 $R_*$ \citep[close to the Alfv\'en radius,][]{Shultz2019} forming auroral rings \citep[see Figures 1 and 2 of][]{Das2020b}. We consider that the intensity drops off as a Gaussian with mean zero and $\sigma$ of $5^\circ$ as the angle between the final ray direction and the line-of-sight increases \citep{Das2020b}. The role of the $\sigma$ parameter is qualitatively similar to that of the beaming angle (that controls the beam-size) in the simulation of \citet{Leto2018}, with a larger beaming angle corresponding to a larger value of the former.

The results of our simulation are shown on the left of Figure \ref{fig:lc_simulation}. The net deflection (defined as the difference between the initial ray direction and the ray direction after exiting the magnetosphere) suffered by the ECME at different frequencies are $3^\circ$, $1.5^\circ$ and $0.06^\circ$ at 700, 1000 and 6000 MHz respectively for the rotational phases corresponding to the maximum observed flux densities. These deviation angles are calculated using the ray paths obtained from the simulation \citep{Das2020b}. Thus, even for these modest deviation angles and relatively small value of $\sigma$ (thus small beam-size), ECME from the star (produced in the southern magnetic hemisphere) is prominent at sub-GHz frequencies and exhibits high circular polarization at the rotational phase where $\langle B_\mathrm{z}\rangle$ is closest to zero (the minimum value of $|\langle B_\mathrm{z}\rangle|$). Note that the data reported by \citet{Leto2018} did not cover this rotational phase. The circular polarization decreases monotonically with increasing observing frequency. 

We next investigate the effect of increasing the value of the $\sigma$ parameter. On the right of Figure \ref{fig:lc_simulation}, we show the lightcurves corresponding to $\sigma=15^\circ$. The primary difference is that in this case, ECME is observable at all rotational phases (even at 6000 MHz), the maximum flux density is observed at the phase where $|\langle B_\mathrm{z}\rangle|$ is closest to zero. The emission has significant circular polarization at sub-GHz frequencies throughout the stellar rotation cycle, but the circular polarization goes to zero at 6000 MHz. These properties closely resemble the observed properties of our target.


In the above simulations, the underlying assumption is that the intrinsic ECME spectrum is flat and identical for both magnetic hemispheres. This prediction, however, depends sensitively on the detail structure of the magnetic field and the magnetosphere. Nevertheless, our simulation provides strong support to the idea that HR\,5907 is another MRP and its emission at uGMRT frequencies is dominated by ECME at all rotational phases.

\subsection{Broad-band Spectra} \label{sec_SED_fit}

The broad-band spectra of HR,5907 exhibit turnovers at both low and high frequencies. These turnovers can be modeled using a combination of different absorption processes (e.g., \citealt{Bloot2022}). For bands with multiple observations, we used the average of all flux densities corresponding to that band. We assume that the spectrum remains approximately flat in the mid-frequency range, spanning from a few GHz to a few tens of GHz. This assumption is reasonable, as evidenced by, for example, the survey of radio emission from magnetic hot stars conducted by \cite{Leto2021}.

\begin{figure}
    \centering
    \includegraphics[width=0.41\textwidth]{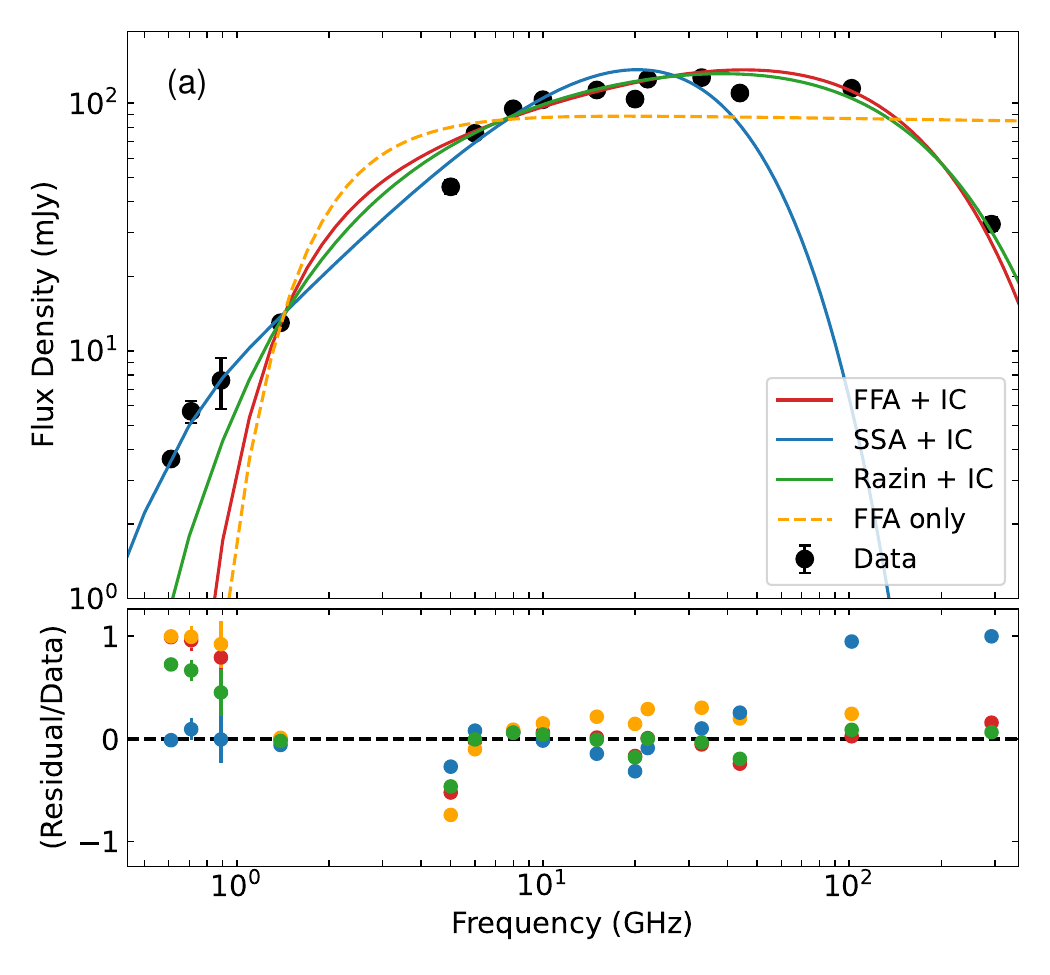}
    \includegraphics[width=0.41\textwidth]{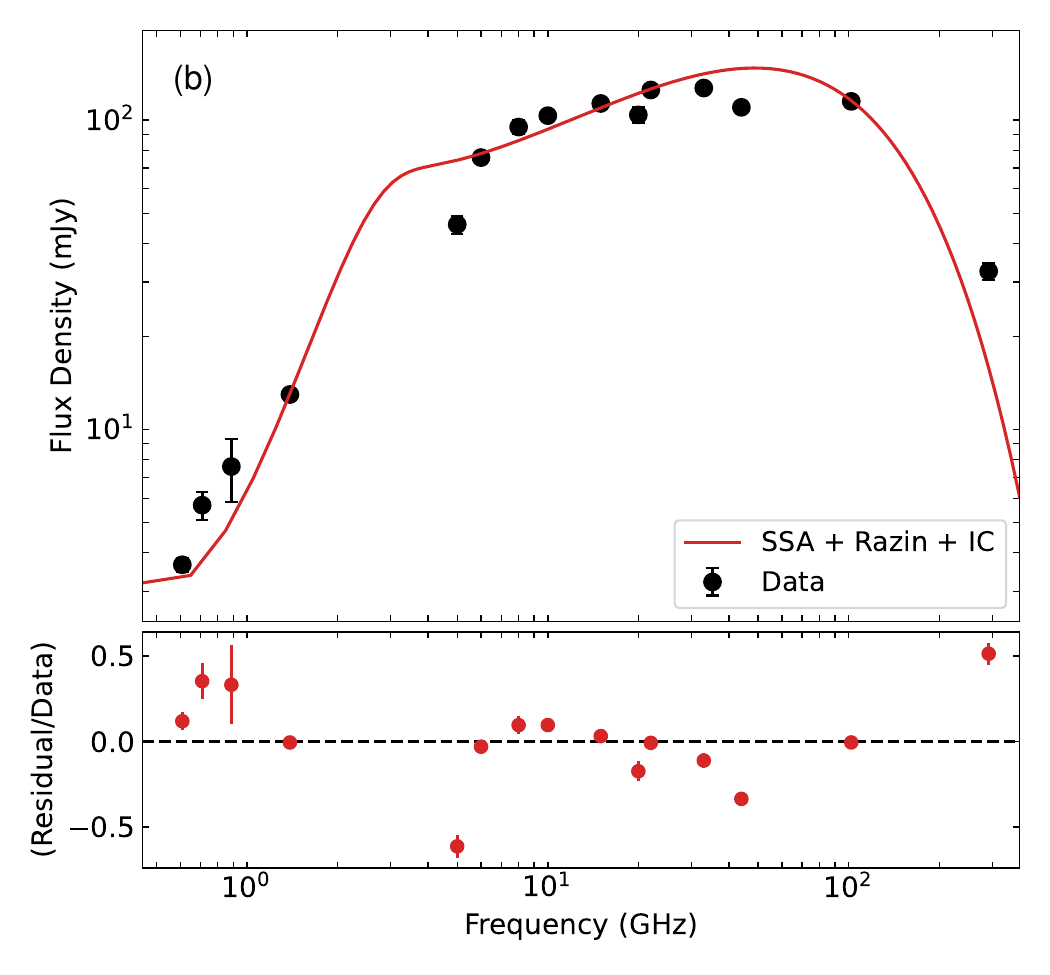}
    \caption{Broad band SED fit of HR\,5907. (a) The top panel shows the 1-component (FFA in yellow) or 2-component (FFA+IC in red; Razin+IC in green; SSA+IC in blue)  fits to the data (in black). In (b), we show the SED fit with a combined SSA+Razin+IC model in red. In both cases, the lower panel indicated the quantity residual/data, with marker colors corresponding to their relevant fitting curves in top panel.  }
    \label{fig:SED_fit}
\end{figure}

To interpret the observed spectral energy distribution (SED), we employ a combination of models incorporating: (1) free-free absorption (FFA), (2) synchrotron self-absorption (SSA), (3) Razin suppression, and (4) inverse Compton (IC) cooling. Details of each model component are discussed in Appendix \ref{Model_Components}. The SED fitting results are presented in Fig. \ref{fig:SED_fit}. Introducing IC cooling is essential to account for the high-frequency turnover. At mid to high frequencies, the Razin and FFA models fit the data best. However, at lower frequencies, the FFA model provides the poorest fit, while the SSA model performs better. If, however, the low-frequency excess is attributed to ECME (electron cyclotron maser emission) that is unaffected by absorption, the FFA model becomes the more physically consistent choice.

We also model the spectra using a combined (SSA+Razin+IC) model, which can be expressed as:

\begin{multline}
      S_\nu = S_{\text{norm}} \left( \frac{\nu}{\nu_{\text{peak,SSA}}} \right)^{-(\beta - 1)/2} \frac{1 - \exp(-\tau_{\text{SSA}})}{\tau_{\text{SSA}}} \\ \times \exp\left( - \frac{\nu_{\text{Razin}}}{\nu} \right) \exp\left( - \frac{\nu}{\nu_{\text{break,IC}}} \right),  
\end{multline}

\noindent where $\alpha$ is the spectral index of the emission in the optically thin regime, $\nu$ is the frequency, $\beta$ is the spectral index related to the energy distribution of the relativistic electrons, $S_{\text{norm}}$ is a normalization constant representing the intrinsic synchrotron flux density, $\nu_{\text{Razin}}$ is the Razin frequency, and $\nu_{\text{break,IC}}$ is the break frequency where the spectrum transitions to the IC-cooled regime. This three-component model provides better results in reproducing the SED than two-component models.

The results indicate the possible presence of background synchrotron emission in the SED. HR,5907's rapid rotation and extremely strong magnetic field generate a large, dense magnetosphere where field-driven high-energy particles from both hemispheres can collide at the equator, producing intense shocks. Another potential origin of the synchrotron emission could be binary interaction. However, no evidence of binarity has been found in the literature for this star. Furthermore, the high-frequency ($>5$ GHz) radio emission from this star aligns with the scaling relationship for gyrosynchrotron emission produced by the CBO mechanism \citep{Shultz2022}.

\section{Conclusion} \label{conclusion}

In this Letter, we investigated the low-frequency radio emission from the rapidly rotating magnetic B star HR\,5907 by observing its full rotation cycle with the uGMRT. The low frequency radio light curve contrasts starkly with previously published high-frequency radio observations \citep{Leto2018}. The radio light curve shows a non-sinusoidal variation with maximum flux near magnetic maximum (which is also the minimum value of $|\langle B_\mathrm{z}\rangle|$). We continuously detect a high fraction of circular polarization ($>20$\%) at all rotational phases, which increases up to $\sim45$\% during the short 3-hour enhancement. The lower limit of brightness temperature was found to be $>10^{12}$ K at all rotation phases. All these emission characteristics point toward a coherent origin of emission at low frequencies. As the star is not known to be in a multiple system, the most compelling explanation is that the observed emission is auroral emission via the electron cyclotron maser emission mechanism. The peak emission of HR\,5907 also falls within the scaling relationship for ECME, as reported by \cite{Das2022scaling}. We thus include HR\,5907 as a new MRP.

We also searched for the possible emission mechanism at higher frequencies and the baseline flux at low frequencies. Based on our findings, we propose the following scenario: The high frequency ($>2$ GHz) emission is dominated by gyrosynchrotron emission, as also suggested by \cite{Leto2018}. The moderate brightness temperature and low fraction of circular polarization ($<5$\%) is consistent with this theory. Also, the position of HR\,5907 in the scaling relation of gyrosynchrotron radio emission at GHz frequencies by \cite{Shultz2022} also supports this assumption. At lower frequencies, the gyrosynchrotron emission is likely to absorbed mostly by free-free absorption. The observed emission at low frequencies is likely entirely dominated by ECME, showing a constant auroral emission during the entire 0.51 d rotation cycle. The geometry of this system is also consistent with the observation of the constant coverage of auroral emission at low frequencies (e.g. see Fig. 3 of \citealt{Leto2020b}). From the SED fit, we suspect additional background contribution from synchrotron emission. However, the explanation for the origin for this synchrotron is not obvious, and will require further study.

\begin{acknowledgments}
We thank the referee for their careful review of our work, and for the very helpful comments.  G.A.W. acknowledges support in the form of a Discovery Grant from the Natural Sciences and Engineering Research Council (NSERC) of Canada. JAB acknowledges support through an NSERC Postgraduate Doctoral Scholarship (PGS D). GH acknowledges the support from the State Research Agency (AEI) of the Spanish Ministry of Science and Innovation and Universities (MCIU) and the European Regional Development Fund (FEDER) under grant PID2021-122397NB-C21/PID2022-136640NB- C22/10.13039/501100011033. The GMRT is run by the National Centre for Radio Astrophysics of the Tata Institute of Fundamental Research. All data exploited in this paper are available from the corresponding public archives. Based on observations made with the Hertzsprung SONG telescope operated on the Spanish Observatorio del Teide on the island of Tenerife by the Aarhus and Copenhagen Universities and by the Instituto de Astrofísica de Canarias. This work has made use of data from the European Space Agency (ESA) mission
{\it Gaia} (\url{https://www.cosmos.esa.int/gaia}), processed by the {\it Gaia} Data Processing and Analysis Consortium (DPAC, \url{https://www.cosmos.esa.int/web/gaia/dpac/consortium}). Funding for the DPAC
has been provided by national institutions, in particular the institutions
participating in the {\it Gaia} Multilateral Agreement.
\end{acknowledgments}

\facilities{uGMRT, Gaia, HIPPARCOS, SONG}

\software{CASA \citep{McMullin2007},
          astropy \citep{Astropy2022},  
          numpy \citep{harris2020array}, 
          Period04 \citep{Lenz2005},
          scipy \citep{Virtanen2020},
          CARTA \citep{Carta2021}
          }

\bibliography{sample631}{}
\bibliographystyle{aasjournal}


\appendix

\section{Stellar parameters} \label{stellar_parameters}

\begin{table}[h!]
    \centering
        \caption{Stellar and magnetospheric parameters of HR\,5907. References: (1) \cite{Hoffleit1991}; (2) \cite{Rivinius2013} (3) \cite{Grunhut2012_hr5907}; (4) \cite{Shultz2019}; (5) \cite{Bailer2021}.}
    \begin{tabular}{lllc}
    \hline \hline
    {Parameters (Units)} & {Symbol} & {Value} & {Ref.} \\
    \hline
    Spectral type & $Sp$ & B2.5V & 1 \\ 
    Eff. temperature (K) & $T_{\rm eff}$ & $17500 \pm 1000$ & 2 \\ 
    Mass $(M_{\odot})$ & $M_*$ & $5.5 \pm 0.5$ & 3  \\ 
    Equatorial radius $(R_{\odot})$ & $R_*$ & $3.1 \pm 0.2$ & 3   \\ 
    Rotational period (d) & $P_{\rm rot}$&  $0.508276^{+0.000015}_{-0.000012}$ & 3 \\ 
    Rot. axis inclination ($^{\circ}$) & $i$ & 70 $\pm 10$ & 3 \\
    Mass-loss rate ($\Dot{M_{\odot}}$ yr$^{-1}$) & $\Dot{M}$ & $-10.10 \pm 0.09$ & 4 \\
    Terminal Velocity (km/s) & $v_{\infty}$ & $1135 \pm 26$ & 4 \\
    Distance (pc) & $d$ & $142 \pm 1$ & 5 \\ 
    $B_d$ from $\langle B_\mathrm{z}\rangle$ (kG) & $B_{\rm d}$ & $15.7^{+0.8}_{-0.9}$ & 3 \\ 
    $B_d$ from Stokes V (kG) & $B_d$ &  $10.4_{-0.4}^{+0.3}$ & 3 \\
    Mag. axis obliquity ($^{\circ}$) & $\beta$ & $7_{-2}^{+1}$ & 3 \\
    Alfv{\'{e}}n radius ($R_{\rm eq}$) & $R_{\rm A}$ & $31 \pm 3$ & 4 \\
    Kepler radius ($R_{\rm eq}$) &  $R_{\rm K}$ & $1.5 \pm 0.1$ & 4 \\ \hline
    \end{tabular}
    \label{tab:my_label}
\end{table}

\section{Updated Ephemeris}\label{Ephemeris_App}


\subsection{Photometric and Spectroscopic Observations}
The \textit{Hipparcos} data set consists of 83 observations obtained between 1990 and 1992 \citep{Hipparcos1997, vanLeeuwen1997}. \cite{Grunhut2012_hr5907} performed a multiharmonic fitting on the \textit{Hipparcos} photometry using a method similar to \cite{Schwarzenberg-Czerny1996}. The authors derived a rotational ephemeris of $\mathrm{HJD}=2447913.694(1) + 0.508276(^{+15}_{-12})\cdot~E$ where $T_{0}$ corresponds to the maximum H$\alpha$ equivalent width (EW). Figure~3 in \cite{Grunhut2012_hr5907} shows that maximum H$\alpha$ EW corresponds to maximum \textit{Hipparcos} magnitude $H_{p}$ and maximum $\langle B_\mathrm{z}\rangle$ (minimum $|\langle B_\mathrm{z}\rangle|$). 

 The phased EW curves and model fits are shown in Figure~\ref{fig:photometry_EW}. The Gaia $G$-band light curve consists of 39 observations obtained between 2014 and 2017 \citep{Gaia2016, Gaia2023}. In our analysis we excluded one outlier with large uncertainty ($\mathrm{BJD}=2457114.95$). To the best of our knowledge the \textit{Gaia} light curve has not been previously analyzed.

We obtained 11 optical spectra of HR~5907 with the robotic 1-m Hertzsprung SONG telescope at Teide Observatory, Tenerife \citep{Andersen_2019}. The SONG observations were obtained over five nights in May 2024 (Program: 24-VAR-02, P.I.: G. Holgado) approximately 100~d before the uGMRT observations. The spectra were acquired with 850~s exposures using slit number 6 which corresponds to a  resolving power of $90,000$. The spectra cover a wavelength range of $4400$ to $6900$\,Å across 51 overlapping orders.  
 
\subsection{Results}
\begin{figure}
    \centering
    \includegraphics[width=0.5\linewidth]{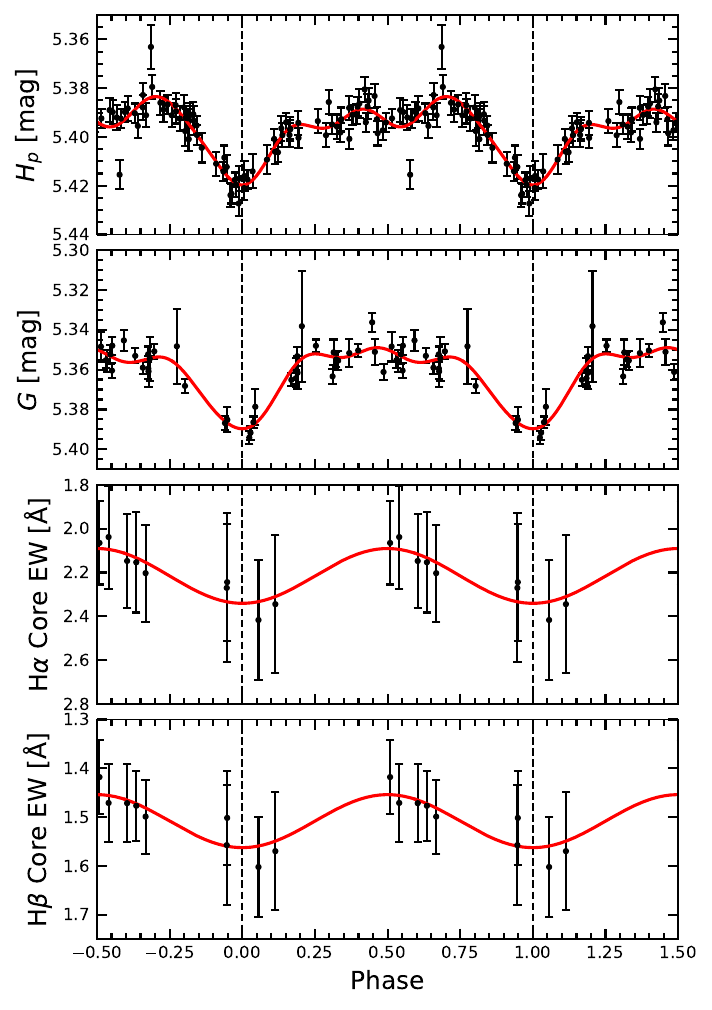}
    \caption{Photometry and EW measurements of HR~5907 as a function of rotation phase. From top to bottom: \textit{Hipparcos} photometry, \textit{Gaia} $G$-band photometry, H$\alpha$ Core EWs and H$\beta$ Core EWs. Each data set is phased to its corresponding ephemeris given in the text.}
    \label{fig:photometry_EW}
\end{figure}

To test the long term stability of the rotation period we performed four-term Fourier fits to the \textit{Hipparcos} and \textit{Gaia} light curves using the Period04 time series analysis software \citep{Lenz2005}. The frequency uncertainties were estimated using the analytical formula derived by \cite{Montgomery1999}. We derived a rotational period of  $P_{H_{p}}=0.508271\pm0.000011$\,d from the \textit{Hipparcos} photometry which is consistent with the period measured by \cite{Grunhut2012_hr5907}. We derived a slightly longer period of $P_{G}=0.508297\pm0.000013$\,d from the \textit{Gaia} photometry. The difference of $\sim2$\,s in the best fitting periods is not significant as the $1.5\sigma$ uncertainty ranges overlap. We take $T_{0}$ as the time of maximum magnitude in the model fit closest to the middle of the observation timespan. We adopt $\mathrm{BJD}^{H_{p}}_{0}=2448370.38632$ and $\mathrm{BJD}_{0}^{G}=2457366.23946$ for \textit{Hipparcos} and \textit{Gaia} respectively. Figure~\ref{fig:photometry_EW} shows the phased light curves with their corresponding Fourier fits. 

We computed EW measurements for the H$\alpha$ and H$\beta$ lines in the SONG spectra, selecting narrow regions around the line cores to reduce noise from the wide profile wings. We computed the EWs over  10\,Å and 5\,Å regions for the H$\alpha$ and H$\beta$ lines respectively. As the SONG observations were obtained over a 5-day period, we cannot compute the current rotation period to high precision. However, because the SONG observations were obtained about 200 rotation cycles before the uGMRT observations, the dominant uncertainty on the inferred rotational phase comes from $T_{0}$, not the period. We performed a joint non-linear least squares sinusoidal fit to the H$\alpha$ and H$\beta$ EW measurements adopting a fixed period of $P=0.5083$\,d. We find $\mathrm{BJD}_{0}^{\mathrm{SONG}}=2460443.684(27)$ with corresponding phase uncertainty $\sigma_{\phi}=0.054$. The phased EW curves and model fits are shown in Figure~\ref{fig:photometry_EW}.

\section{Log of new Radio Observations}

The flux densities and circular polarization fraction obtained from the new uGMRT observations are given in Table \ref{tab:gmrtflux}.

\begin{deluxetable}{cccccc}
\tablecaption{Variation of the flux density of HR 5907 with rotational phase.   \label{tab:gmrtflux}}
\tablehead{
\colhead{Mean HJD}  & \colhead{Mean Phase} &  \multicolumn{3}{c}{Band 4} & \colhead{Band 5} \\ 
\colhead{}  & \colhead{(SONG ephem.)} & \colhead{RR flux} & \colhead{LL flux} & \colhead{Polarization} & \colhead{Stokes I Flux Density}  \\
\colhead{} & \colhead{} & \colhead{(mJy)} & \colhead{(mJy)} & \colhead{Fraction (\%)}  & \colhead{(mJy)} 
}
\startdata
\multicolumn{6}{c}{Observation 1 (1 August 2024)} \\
2460524.98335 & 0.94 & 6.99 $\pm$ 0.40 & 4.03 $\pm$ 0.34 & 32 $\pm$ 6 & 7.87 $\pm$ 0.63 \\
2460525.01252 & 1.00 & 9.70 $\pm$ 0.48 & 3.87 $\pm$ 0.20 & 55 $\pm$ 6 & 7.22 $\pm$ 0.39 \\
2460525.0389 & 1.05 & 14.10 $\pm$ 1.40 & 5.55 $\pm$ 0.66 & 46 $\pm$ 10 & 11.80 $\pm$ 1.00\\
2460525.06807 & 1.11 & 9.29 $\pm$ 0.92 & 5.63 $\pm$ 0.63 & 24 $\pm$ 8 & 12.70 $\pm$ 1.20\\
2460525.09584 & 1.16 & 7.11 $\pm$ 0.58 & 5.35 $\pm$ 0.42 & 14 $\pm$ 6 & 10.25 $\pm$ 0.73\\
2460525.12362 & 1.22 & 7.23 $\pm$ 0.44 & 4.37 $\pm$ 0.46 & 25 $\pm$ 6 & 7.96 $\pm$ 0.79\\
2460525.15139 & 1.27 & 7.47 $\pm$ 0.49 & 5.19 $\pm$ 0.37 & 18 $\pm$ 5 & 8.47 $\pm$ 0.74\\
2460525.19653 & 1.36 & 8.09 $\pm$ 0.71 & 6.03 $\pm$ 0.89 & 16 $\pm$ 9 & 4.54 $\pm$ 0.25\\
2460525.22083 & 1.41 & 8.24 $\pm$ 0.73 & 5.71 $\pm$ 0.66 & 20 $\pm$ 8 & 7.53 $\pm$ 0.68\\
\multicolumn{6}{c}{Observation 2 (15 August 2024)} \\
2460538.93273 & 1.39 & 6.38 $\pm$ 0.36 & 3.75 $\pm$ 0.29 & 26 $\pm$ 5 & - \\
2460538.95703 & 1.43 & 5.18 $\pm$ 0.51 & 2.75 $\pm$ 0.42 & 31 $\pm$ 9 & - \\
2460538.98133 & 1.48 & 7.79 $\pm$ 0.25 & 4.00 $\pm$ 0.12 & 32 $\pm$ 3 & - \\
2460539.00564 & 0.53 & 6.39 $\pm$ 0.27 & 4.07 $\pm$ 0.20 & 22 $\pm$ 3 & - \\
2460539.03133 & 0.58 & 8.49 $\pm$ 0.44 & 4.53 $\pm$ 0.35 & 30 $\pm$ 5 & - \\
2460539.05494 & 0.63 & 6.31 $\pm$ 0.22 & 4.44 $\pm$ 0.22 & 17 $\pm$ 3 & - \\
2460539.07924 & 0.68 & 6.37 $\pm$ 0.47 & 3.91 $\pm$ 0.37 & 24 $\pm$ 6 & - \\
2460539.10354 & 0.72 & 5.68 $\pm$ 0.40 & 2.84 $\pm$ 0.30 & 33 $\pm$ 6 & - \\
2460539.13965 & 0.79 & 4.49 $\pm$ 0.26 & 2.29 $\pm$ 0.28 & 29 $\pm$ 5 & - \\
2460539.16673 & 0.85 & 3.41 $\pm$ 0.34 & 1.43 $\pm$ 0.18 & 42 $\pm$ 9 & - 
\enddata
\end{deluxetable}

\section{Components of Spectral Modeling} \label{Model_Components}

\subsection{Free-free absorption}

{Free-free absorption (FFA) occurs when ionized gas surrounding the synchrotron source absorbs the emitted radiation. The absorbing medium, consisting of free electrons and ions, becomes increasingly opaque at lower frequencies, causing an exponential turnover. The optical depth of FFA, $\tau_{\text{FFA}}$, can be expressed as \citep{Mezger1967}:

\begin{equation}
    \tau_{\text{FFA}} \approx 8.24 \times 10^{-2} \nu^{-2.1} T_e^{-1.35} \int n_e^2 dl,
\end{equation}

\noindent
where $\nu$ is the frequency, $T_e$ is the electron temperature, and $n_e$ is the electron density along the line of sight. Particularly for massive stars, the free-free optical depth can be re-written as \citep{Torres2011}:

\begin{equation}
    \tau_{\rm ff} = 5 \times 10^3 \  \dot{M}_{-8}^2 V_{\infty}^{-1}  \nu_{\rm GHz}^{-2} T_{\rm wind}^{-3/2} D_{\rm ff}^{-3},
\end{equation}

\noindent
where $D_{\rm ff}$ is the distance from the star in units of $3 \times 10^{12}$ cm, and $\dot{M}_{-8}$ is the mass-loss rate (in units of $10^{-8} M_{\odot}/{\rm yr}$). The flux density under FFA is modeled as:

\begin{equation}
    S_\nu = S_{\text{norm}} \left( \frac{\nu}{\nu_{\text{peak,FFA}}} \right)^{\alpha} \exp\left( - \tau_{\text{FFA}} \right),
\end{equation}

\noindent
where $\nu_{\text{peak,FFA}}$ is the frequency at which the optical depth $\tau_{\text{FFA}}$ reaches unity, and $\alpha$ is the spectral index of the synchrotron emission in the optically thin regime. 
}

\subsection{Synchrotron self-absorption}

{Synchrotron self-absorption (SSA) occurs when relativistic electrons that emit synchrotron radiation reabsorb the emitted photons. The effect is most pronounced at lower frequencies where the emitted photons are less energetic, resulting in a suppression of the observed flux density below a critical frequency, known as the peak frequency $\nu_{\text{peak,SSA}}$. The SSA process introduces a characteristic turnover in the radio spectrum, with the following spectral shape for the flux density $S_\nu$ \citep{Tingay2003}:

\begin{equation}
    S_\nu = S_{\text{norm}} \left( \frac{\nu}{\nu_{\text{peak,SSA}}} \right)^{-(\beta - 1)/2} \frac{1 - \exp(-\tau_{\text{SSA}})}{\tau_{\text{SSA}}},
\end{equation}

\noindent
where the optical depth $\tau_{\text{SSA}}$ is given by:

\begin{equation}
    \tau_{\text{SSA}} = \left( \frac{\nu}{\nu_{\text{peak,SSA}}} \right)^{-(\beta + 4)/2}.
\end{equation}

Here, $\nu$ is the frequency, $\beta$ is the spectral index related to the energy distribution of the relativistic electrons, and $S_{\text{norm}}$ is a normalization constant representing the intrinsic synchrotron flux density.}

\subsection{Razin effect}

The Razin effect, or the Tsytovich-Razin effect, occurs when the synchrotron emission is suppressed by the presence of a dense plasma. The effect causes a reduction in the effective magnetic field experienced by relativistic electrons, leading to a suppression of synchrotron radiation at frequencies below the Razin frequency $\nu_{\text{Razin}}$ (e.g. \citealt{Rybicki1979, Dougherty2003, Erba2022}). The spectrum for the Razin effect is modeled by an exponential suppression term :

\begin{equation}
    S_\nu = S_{\text{norm}} \nu^\alpha \exp\left( - \frac{\nu_{\text{Razin}}}{\nu} \right),
\end{equation}

\noindent
where $\alpha$ is the spectral index of the synchrotron emission in the optically thin regime. The Razin frequency ($\nu_{\rm Razin}$) given by \cite{Ginzburg1965}:

\begin{equation}
    \nu_{\rm Razin} = 2 \times 10^{-8} \frac{n_{\rm e}}{B} \ \ {\rm GHz},
\end{equation}

\noindent
where $B$ is the magnetic ﬁeld strength in G, and $n_{\rm e}$ is the electron density in cm$^{-3}$.

\subsection{Inverse-Compton Cooling}

  {At high frequencies, inverse-Compton (IC) cooling becomes the dominant process for energy loss in relativistic electrons. In this process, photons from the star's radiation field scatter off relativistic electrons, gaining energy and thereby cooling the electron population. This cooling effect manifests as a high-frequency cut-off in the observed radio spectrum. The flux density in the presence of IC cooling is modeled as \citep{Kellermann1969, Komissarov1994}:

\begin{equation}
    S_\nu = S_{\text{norm}} \nu^\alpha \exp\left( - \frac{\nu}{\nu_{\text{break,IC}}} \right),
\end{equation}

\noindent
where $\nu_{\text{break,IC}}$ is the break frequency at which the spectrum transitions from the synchrotron-dominated regime to the IC-cooled regime.}

\end{document}